\documentclass[%
 reprint,
 amsmath,amssymb,
 aps,
 prl,
]{revtex4-1}
\usepackage{hyperref}
\usepackage{bbold}
\usepackage{graphicx}
\usepackage{dcolumn}
\usepackage{bm}

\usepackage{braket}
\usepackage{xcolor}
\usepackage{csquotes}

\usepackage{changes}

\begin{document}

\preprint{APS/123-QED}

\title{Extended local ergotropy}

\author{Riccardo Castellano$^{1,2,3,*}$, Donato Farina$^{2,4,\dag}$, Vittorio Giovannetti$^{5}$, Antonio Acin$^{2,6}$}
\affiliation{${}^{1}$Scuola Normale Superiore, I-56126 Pisa, Italy}
\affiliation{${}^{2}$ICFO - Institut de Ciencies Fotoniques, The Barcelona Institute of Science and Technology, 08860 Castelldefels, Barcelona, Spain}
\affiliation{${}^{3}$Dipartimento di Fisica dell’Università di Pisa, Largo Pontecorvo 3, I-56127 Pisa, Italy}
\affiliation{${}^{4}$Physics Department E. Pancini - Università degli Studi di Napoli Federico II,
Complesso Universitario Monte S. Angelo - Via Cintia - I-80126 Napoli, Italy}
\affiliation{${}^{5}$NEST, Scuola Normale Superiore and Istituto Nanoscienze-CNR, I-56127 Pisa, Italy}
\affiliation{${}^{6}$ICREA - Institució Catalana de Recerca i Estudis Avan\c cats, 08010 Barcelona, Spain
\\
$^{*}$riccardo.castellano@sns.it, $^{\dag}$donato.farina@unina.it}


\date{\today}

\begin{abstract}
{A fundamental problem in quantum thermodynamics is to properly quantify the work extractable from out-of-equilibrium systems. While for closed systems, maximum quantum work extraction is defined in terms of the ergotropy functional, this question is unclear in open systems interacting with an environment. The concept of local ergotropy has been proposed, but it presents several problems, such as 
it is not guaranteed to be non-increasing in time.}
Here we introduce the concept of \textit{extended local ergotropy} by exploiting the free evolution of the system-environment compound.
At variance with the local ergotropy, the extended local ergotropy is greater, is non-increasing in time, and activates the potential of work extraction in many cases.
We then concentrate on specific schemes in which we alternate repeated local unitaries and free system-environment evolution.
We provide examples based on the Jaynes-Cummings model, presenting practical protocols and analytic results that serve as proof of principle for the aforementioned advantages.

\end{abstract}

\maketitle


{How much work can be extracted from an out-of-equilibrium quantum system?}
For an isolated quantum system S, described by Hamiltonian $H_{\rm S}$ and state $\rho_{\rm S}$, the Ergotropy functional \cite{allahverdyan2004maximal} is commonly accepted as the figure of merit for the maximum work extractable under cyclic protocols \cite{GibbsFormula,Ergotropy,ErgotropyAut,ErgotropyEM}. It is defined as
 \begin{equation}
{\mathcal{E}}(\rho_{\rm S},H_{\rm S}):=\underset{U_{\rm S} \in \mathcal{U}_{\rm S}}{\max}\;\; {\rm tr}[H_{\rm S}(\rho_{\rm S}-U_{\rm S}\rho_{\rm S} {U_{\rm S}}^\dag)]\,,
\label{ergo}
\end{equation}
where $\mathcal{U}_{\rm S}$ is the set of unitaries on S.
Remarkably, the optimal unitary has a closed expression in terms of the eigenvectors of $H_{\rm S}$ and $\rho_{\rm S}$ \cite{ErgoFormula}.\\
However, in practical experimental settings, the system of interest S interacts (weakly or not) with external degrees of freedom, the environment E, on which we typically have very limited control.
Hence, it becomes crucial to properly include the environment E and its interaction with the system S in a work extraction task.
Regarding this, a series of approaches have been proposed so far.
Ergotropy Extraction (EE) via thermal operations has been proposed in Ref.\,\cite{ThermalOp}. Thermal operations mimic classical interaction between S and E, i.e. they keep unchanged the sum of the local energies of S and E \cite{Faist_2015}. This construction, however, requires in general a detailed engineering of the S-E interaction. 
One can instead consider all possible {quantum channels}, defined by Completely Positive Trace Preserving (CPTP) maps, acting locally on S as allowed operations \cite{StrongLocal}. 
Consequently, states whose energy cannot be decreased by any local CPTP map are named local CP-passive, and semi-definite programming techniques {can be} used  to characterize them and compute upper bounds on the extractable energy~\cite{LocalEnergyExtractionCPTP}.

However, in this case, the energy extracted is not unequivocally accepted as work as it typically implies an entropy change.
More recently, in Ref.\,\cite{LErgo}, and in analogy with ergotropy, the set of allowed operations is restricted to all unitaries acting locally on the S sub-system. 
Such notion of local ergotropy (LE) is defined as
 \begin{equation}
    {\mathcal{E}}_{\rm S}(\rho_{\rm SE},H_{\rm SE}):=\underset{U_{\rm S} \in \mathcal{U}_{\rm S}}{\max}\;\; {\rm tr}[H_{\rm SE}(\rho_{\rm SE}-U_{\rm S}\rho_{\rm SE}U_{\rm S}^\dag)]
    \label{localErgo}\,,
\end{equation}
where $\rho_{\rm SE}$ is the joint state of SE, $ H_{\rm SE}$ is the full interacting Hamiltonian
\begin{equation}
    H_{\rm SE}=H_{\rm S}+H_{\rm E}+V_{\rm SE}\,,
    \label{Hse}
\end{equation}
with local terms $H_{\rm S}$, $H_{\rm E}$ and interaction $V_{\rm SE}$.
Eq.\,\eqref{localErgo} expresses the fact that it is not possible to control the environmental degrees of freedom and the interaction between the systems. 
Nonetheless, the energy functional is evaluated with respect to the full interacting Hamiltonian $ H_{\rm SE}$ making SE correlations a resource in several cases.\\
Both the definitions introduced in \cite{StrongLocal,LErgo}, namely CP-passivity and LE, share a notable operational problem: they are not guaranteed to be non-increasing in time.
As the time evolution of the joint system is not local, both
{quantities} are time-dependent when the state is evolving under the compound Hamiltonian $H_{\rm SE} $.
{This leads to exotic situations, such as having passive states \cite{allahverdyan2004maximal} according to these quantities (that is states with zero extractable work) evolving into non-passive ones by the natural SE dynamics with no active intervention on S (see, e.g., Fig.~1 of Ref.~\cite{LErgo})}.\\
A second important element to take into account is the adiabatic approximation underlying the definitions {of CP-passivity or LE.}
In order to eventually generate a local map on S, it is necessary to assume that the local manipulation is carried out on a much shorter time scale than the one {governing} the dynamics of the joint compound SE.

The objective of this {work} is to introduce the concept of Extended Local Ergotropy (ELE), a figure of merit {for work extraction in open systems} that solves {all the previous} issues: {(i) it is non-increasing under the natural SE dynamics, (ii) it does not require any adiabatic approximation and (iii) it does not involve any} control on the environmental degrees of freedom.
{The main point behind our construction is to take into account} the free evolution of the SE compound and its entangling potential to enlarge the set of implementable {unitary operations}.\\

{\it Extended Local Ergotropy.---}
To start, we define the set $\bar {\mathcal{U}}_{\rm ex}(H_{\rm SE}) $ (or $\bar{\mathcal{U}}_{\rm ex}$ for compactness) of Extended Local Unitaries as the closure \cite{Closure}
of the set
\begin{equation} \label{LocalOp}
    \mathcal{U}_{\rm ex}(H_{\rm SE}):=\big\{ U\; s.t.\; U = \mathcal{T} \exp \big [-i\int_{0}^{t_f} \!\!\! H_{\rm SE}+H_{\rm S}(t)dt \big ]\big \} \,.
\end{equation}
{Here, we do not allow for Hamiltonian control of E, but only on S.}
These unitaries {have a clear operational character: they} are not strictly local on S, {but the nonlocality is provided only by the natural SE dynamics, in particular} by its interacting term $V_{\rm SE}$.
ELE then reads \cite{cyclic}
\begin{equation}
    {\mathcal{E}}_{\rm ex}(\rho_{\rm SE},H_{\rm SE}):= \underset{U \in \bar {\mathcal{U}}_{\rm ex} }{\rm sup}  {\rm tr}[H_{\rm SE}(\rho_{\rm SE}-U \rho_{\rm SE} U^\dag)]
    \label{LEE}
    \,.
\end{equation}
When the SE {system} is finite-dimensional, the $ \rm sup$ can be replaced by a $\rm max $.
We notice that, by construction, 
 ELE is always greater or equal than LE 
 {and smaller or equal than the Global Ergotropy (GE) ${\cal E}(\rho_{SE}, H_{SE})$  of the  $SE$ compound which measures the work extractable when global operations are 
granted in the model, i.e.}
 \begin{equation}
     {\mathcal{E}}_{\rm S}(\rho_{\rm SE},H_{\rm SE})  \leq {\mathcal{E}}_{\rm ex}(\rho_{\rm SE},H_{\rm SE}) \leq {\mathcal{E}}(\rho_{\rm SE},H_{\rm SE})\,.  \label{G=ExL>L}
 \end{equation}
Also, the functional \eqref{LEE} is convex in the state $\rho_{\rm SE}$ and, in contrast with local ergotropy, 
{non-increasing under time evolution induced by the free Hamiltonian $H_{\rm SE}$,}
\begin{equation}
     {\mathcal{E}}_{\rm ex}(\rho_{\rm SE}(t),H_{\rm SE}) \leq {\mathcal{E}}_{\rm ex}(\rho_{\rm SE}(0),H_{\rm SE}) \,,\quad \forall t \geq 0.
     \label{decreasing}
\end{equation}
Indeed, since time evolution is part of the $\bar{\mathcal{U}}_{\rm ex} $ set, ELE can only decrease in time, proving \eqref{decreasing}.
{This is a property generically demanded to a quantum resource. For instance, entanglement is non-increasing under LOCC operations. Analogously, ELE is non-increasing under free SE evolution.}
{ Notably}, the equality 
 \begin{equation}
     {\mathcal{E}}_{\rm ex}(\rho_{\rm SE}(t),H_{\rm SE})={\mathcal{E}}_{\rm ex}(\rho_{\rm SE}(0),H_{\rm SE})  \,,\quad \forall t \geq 0\,,
 \label{constant}
 \end{equation}
holds if the $H_{\rm SE} $ is bounded and has discrete spectrum.
The proof \cite{supp-mat} relies on the quasi-recurrence of unitary evolutions \cite{Q.Recurr,Noteonrecurr}.
In \cite{Recurr} estimates and upper bounds for such recurrence time are reported. 
Nevertheless, these values are excessively long to be {relevant} in {any realistic scenario}. Consequently, in {any practical situation}, one should consider ELE as a quantity that {strictly} decreases over time.  Interestingly, it can {also} be shown \cite{supp-mat} that a continuous Hamiltonian allows for irreversible flows of energy from S to E, making ELE strictly decreasing in time.  

{\it Bang-bang representation.---}
{
Bang-Bang control procedures are obtained by abruptly alternating between two different types of Hamiltonian drivings.
Such evolutions proved to be very effective in generating optimal control pulses in different contexts of quantum information \cite{PhysRevLett.82.2417, morton2006bang}.
In our case we can show that, as long as the SE compound is finite-dimensional, all possible elements of $\bar{\cal U}_{\rm ex}$  
can be generated via bang-bang sequences formed by free evolutions of the system and by strong driving pulses on S \cite{supp-mat}.
Indeed, if the dimension of the Hilbert of $\rm SE$ is finite,  Quantum Control Theory assesses that $\bar{\cal U}_{\rm ex}$  is a compact and connected Lie group given by the exponential of the dynamical Lie algebra \cite{Q.ControlBook,Q.ControlRev}.
Accordingly it follows that  given  $U\in \bar{\cal U}_{\rm ex}$   it can be expressed as 
\begin{equation} \label{bang-bang}
U = U_{\rm S}^{(\mathcal{N}-1)}U_{0}(\delta t_{\mathcal{N}-1}) 
\dots
{U_{\rm S}^{(0)}}U_{0}(\delta t_0)\,,
 \end{equation} 
obtained by alternating unitary operations on the system, $U_{\rm S}^{(0)},\cdots,U_{\rm S}^{(\mathcal{N}-1)} \in {\cal U}_S$  with time intervals $\delta t$ 
of free-time evolution, defined by the operator $U_0(\delta t) := \exp(-i\delta t H_{SE})$. Further, the number of necessary unitaries ${\cal  N}$ is uniformly bounded. 
This result does not always apply to infinite-dimensional models. 
For completeness, however, as an example of the treatment in the infinite-dimensional discrete case, we will analyse the Jaynes-Cummings model. Systems with continuous spectra, instead, warrant a separate analysis that goes beyond the scope of this work.
}

{{{\it Saturation of the global entropy.---} While under general conditions the gap between  ELE and GE is expected to be a strict one, for some models the two values may coincide.
An example is provided by a 1D chain of $M$ $1/2$-spins with 
Heisenberg-like nearest neighbour interaction,
\begin{eqnarray}
   H_{SE}=\underset{n={1}}{\overset{M-1}{\sum}} \gamma (\sigma^{(n)}_{x}\sigma^{(n+1)}_{x}+\sigma^{(n)}_{y}\sigma^{(n+1)}_{y}+\Delta \sigma^{(n)}_{z}\sigma^{(n+1)}_{z})\,,
   \nonumber
\end{eqnarray}
with $\sigma_{x,y,z}^{(i)}$ describing the Pauli operators of the $i$-th chain element. In this case, identifying  the first spin element  with $\rm S$ and the remaining ones with  the environment $\rm E$, using the results of quantum control \cite{LocalControlSpinChain(General)}, one can show that
  $\mathcal{E}_{\rm ex}(\rho_{\rm SE},H_{\rm SE})= \mathcal{E}(\rho_{\rm SE},H_{\rm SE})$
via appropriate manipulations of the local magnetic field acting on $\rm S$  \cite{supp-mat}. Interestingly, for $\Delta=0$, controlling instead the first two spins even allows efficient control (i.e., quadratically longer than for direct
control) \cite{ControlSpinChain-2spins}.

A further example of a system for which ELE and GE can coincide is represented by the case where $S$ is a two-level {(qubit)} system
interacting with a single electromagnetic cavity mode via a Jaynes-Cummings (JC) interaction \cite{J-C-Original}. In this case the Hamiltonian
of the SE compound is given by (setting $\hbar=1$) 
\begin{equation}
H_{\rm JC}:=\omega_{\rm S}\frac{\sigma_z+\mathbb{1}}{2} +\omega_{\rm E}a^{\dagger}a +\frac{\Omega}{2}(\sigma^{+}\otimes a +\sigma^{-} \otimes a^{\dagger})
\,,
\end{equation}
where 
$a^\dag$ ($a$) is the bosonic creation (annihilation) operator of the cavity mode,
{\color{black}$ \omega_{\rm S}$ and $  \omega_{\rm E}$ are, respectively, the atom's gap energy and the cavity's frequency, and 
$\sigma^{\pm}:=({\sigma_{x}\pm i\sigma_{y}})/{2} $.
Let $\Delta\omega:=\omega_{\rm S} - \omega_{\rm E} $ be the detuning and $\phi_{n}:=\frac{1}{2}\arctan{({\sqrt{n+1}\, \Omega}/{\Delta\omega} )} $.  
For  $\sqrt{\Delta \omega^2 +\Omega^2} < (\omega_{\rm E}+\omega_{\rm S})$,
the Hamiltonian's ground state is $\ket{00}_{\rm SE}$ with eigenvalue $E_{0}=0$,
where $\ket{0}_{\rm S}$ and $\ket{1}_{\rm S}$ are the eigenstates of $\sigma_z$ corresponding to eigenvalues $-1$ and $1$, and $\{\ket{n}_{\rm E}\}_{n \in \mathbb{N}}$ are bosonic Fock states. 
The excited states are
\begin{eqnarray}
\nonumber
\ket{n+}=\cos\phi_{n}\ket{1}_{\rm S}\otimes\ket{n}_{\rm E} +\sin{\phi_{n}}\ket{0}_{\rm S}\otimes\ket{n+1}_{\rm E} \\
\ket{n-}=\sin\phi_{n}\ket{1}_{\rm S}\otimes\ket{n}_{\rm E} -\cos{\phi_{n}} \ket{0}_{\rm S}\otimes\ket{n+1}_{\rm E}
\nonumber
\\
\end{eqnarray}
with corresponding eigenvalues 
\begin{eqnarray}
E_{n,\pm}=\omega_{\rm S}/2+\omega_{\rm E}(n+1/2) \pm \Delta \omega_n\,,
\end{eqnarray}
and
$
\Delta \omega_n :=\frac{1}{2}\sqrt{\Delta\omega^2 + (n+1)\Omega^2}$.}
For this model, we have shown \cite{supp-mat} that local controls on the spin degrees of freedom are enough to have approximate density matrix controllability on the  
{\color{black}joint system. More precisely,
 for each pair of unitarily equivalent states $\rho_{\rm SE}, \sigma_{\rm SE} $, $\forall \; \epsilon>0 $ and for almost-all values of the coupling constant $\Omega $ \cite{ Q.Control(1972),IndirectControl,AlgConditionsIndirectControll,J-Controllab,InfDimControllabTest}:
\begin{equation}
\exists \;\; U \in \mathcal{U}_{\rm ex}(H_{\rm JC}) \;\; {\rm s.t.} \;\; {\rm tr}[(U\rho_{\rm SE}U^{\dagger}-\sigma_{\rm SE})^2]\leq \epsilon
\,.
\end{equation}
This means we can always bring the initial state arbitrarily close to its passive state $\rho_{\rm SE}^{\rm P}$. However, since $H_{\rm JC} $ is unbounded this does not guarantee that the energy of the final state can be brought arbitrarily close to ${\rm tr}[H_{\rm JC}\rho_{\rm SE}^{\rm P}]$.}
Despite this, for any finite-dimensional approximation of  the the compound, we have
$\mathcal{E}(\rho_{\rm SE},H^{(N)}_{\rm SE})= \mathcal{E}_{\rm ex}(\rho_{\rm SE},H^{(N)}_{\rm SE})$, 
where $H^{(N)}_{\rm SE}$ is a truncation of $H_{\rm JC}$ obtained by only allowing up to $N$ photons  in the cavity. 
However, it is not clear how to construct such a unitary, if there exist upper bounds on the number of operations or on the time needed to implement it. For this reason, in the next section we show for different classes of initial states the amount of work that can be extracted with minimal protocols, involving a small number of operations. }}

\begin{figure}[h]
\centering    \includegraphics[width=0.85\linewidth]{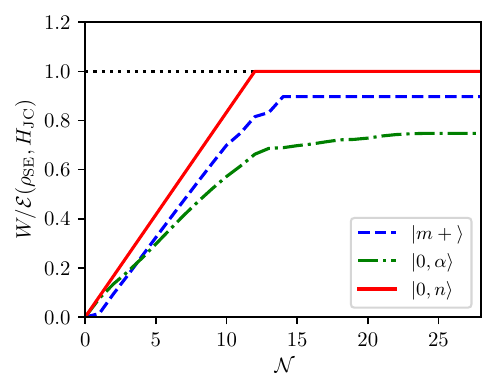}
\caption{
{{Lower bound $W$ for extended local ergotropy in the case of the Jaynes-Cummings model, obtained using a slight generalisation of the bang-bang unitary of Eq.\,\eqref{bb-prot} (see \cite{supp-mat} for details).}}
We plot the ratio between $W$ and the GE, as a function of the number of steps $\mathcal{N}$. We consider different input states: $\ket{0,n}$, qubit in the ground-state and cavity in a Fock state (red); $\ket{0,\alpha}$, qubit in the ground-state and cavity in a coherent state (green dash-dotted); eigenstate $\ket{m+}$ of the Hamiltonian of SE (blue dashed).
Initial resources are such that 
initial states have approximately the same GE \cite{AlmostsameGE}.
With input $\ket{0,n}$
the bang-bang protocol is optimal even in saturating the GE bound.
We have set $n=\vert\alpha\vert^2=12$, $m=11 $,
$\omega_{\rm S}= \omega_{\rm E}=1$, $\Omega=0.1$.
}
\label{Fig.1}
\end{figure}

{\it Practical protocols for the JC model.---}
{{As a first example, we consider the case where the two-level system of the JC model is in the ground state $|0\rangle$ and the cavity mode is in a Fock state $|n+1\rangle$.
}}
%
Here, for zero detuning $ \Delta \omega:= 0$, we can alternate work extraction unitaries on the system (bit flips) with free time evolutions to re-excite the atom and extract one by one all the photons stored in the cavity. This makes the bang-bang protocol saturating the ELE and the GE of SE.
Let $\delta t_{n}:=\frac{\pi}{2\Delta \omega_{n}} $ and $ U^{\rm (bf)}_{\rm S}$ be the bit-flip on qubit S, then \cite{supp-mat}
\begin{equation}
\label{bb-prot}
  U^{\rm (bf)}_{\rm S}U_{0}(\delta t_0)U^{\rm (bf)}_{\rm S}...  U^{\rm (bf)}_{\rm S} U_{0}(\delta t_{n})\ket{\psi_{\rm in}}_{\rm SE}=\ket{00}_{\rm SE}\,.
\end{equation}
This implies
 \begin{equation}
     {\mathcal{E}}_{\rm ex}(\ket{0}_{\rm S}\otimes\ket{n+1}_{\rm E})={\mathcal{E}}(\ket{0}_{\rm S}\otimes\ket{n+1}_{\rm E})=(n+1) \omega_{\rm S}\,,
 \end{equation}
proving that ELE can saturate the ergotropy of SE. 
We remark that, instead, for any $n$ and value of the detuning,  {the LE is zero}, ${\mathcal{E}}_{\rm S}(\ket{0}_{\rm S}\ket{n}_{\rm E})=0 $. All this is depicted in Fig.\,(\ref{Fig.1}) where 
we plot the work extracted in terms of the number of steps $\mathcal{N}$ in the bang-bang protocol for input state $\ket{0,n}$.
ELE reaches the GE value for $\mathcal{N}=n$.
For completeness, we report work extracted with a generalization of the protocol described in Eq.~(\ref{bb-prot}) \cite{supp-mat} for different input states.
We consider  
qubit in the ground state and cavity in a coherent state, $\ket{0, \alpha}$, and the eigenstate of the Hamiltonian of SE, $\ket{m+}$,
with initial parameters ($n$, $\alpha$ and $m$) such that 
initial states have approximately the same GE \cite{AlmostsameGE}. 
Also for these two examples, the final value of the ELE is considerably high, despite not reaching GE performances. 
\\

{\it Thermal states.---}
We now move on to the case in which the qubit and the cavity {{of the JC model}} are in a tensor product of two thermal states. We then set
\begin{equation}
\label{thermal-input}
    \rho_{\rm in}=\gamma_{\beta_{\rm S}}\otimes \gamma_{\beta_{\rm E}} \,,
\end{equation}
where the two thermal states refer to the respective local Hamiltonians and $\beta_{\rm X}:=1/(k_{\rm B}T_{\rm X})$ are the inverse temperatures. 
First of all, as for the Fock initial state, we have zero LE, ${\mathcal{E}}_{\rm S}(\gamma_{\beta_{\rm S}}\otimes \gamma_{\beta_{\rm E}},H_{\rm SE})=0 $ \cite{LE=0}.
This means that any observed nonzero ELE implies an advantage with respect to LE.
In this context it is also natural to compare ELE performances with respect to the ergotropic extraction framework defined in Ref.\,\cite{ThermalOp} which is based on
thermal operations.

In order to fairly make comparisons with the latter, we will suppose that we can neglect the interaction when evaluating the energy functionals, this makes the energy functional depending only on the local states. However, this does not mean that the existence of the interaction is not important. Indeed, the set of allowed unitary operations $\mathcal{U}_{\rm ex}(H_{\rm SE})$  is independent of the strength of the interaction \cite{InteractionStrengh}.
Moreover, it is important to note that there is, in general, no trivial ordering between the optimal protocols of the ELE and EE settings, because the set of unitary energy-preserving maps on SE is neither bigger nor smaller than 
$ \bar {\mathcal{U}}_{\rm ex}(H_{\rm SE})$.

\begin{figure}[h] 
    \centering
    \includegraphics[width=0.85\linewidth]{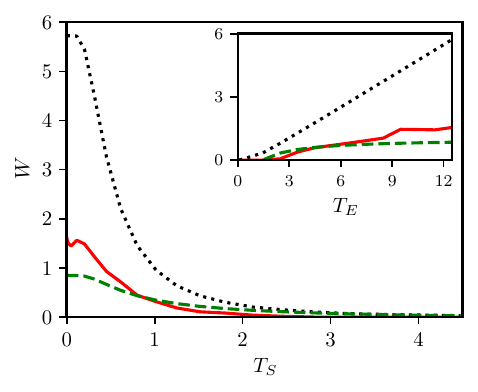}    
\caption{
Work $W$ extracted via $U_{\rm bb}(T_{\rm S},T_{\rm E})$ (red) and maximum Extractable Ergotropy, as defined in Eq.(52) of Ref.\cite{ThermalOp}, (green dashed) for a qubit at temperature $T_{\rm S}$ and bath at temperature $T_{\rm E}$. 
{Both curves are computed for varying $T_{\rm}$,} while we fixed $T_{\rm E}$ to the value corresponding to $12$ average photons.
Inset: 
work extracted via $U_{\rm bb}(0,T_{\rm E})$ (red)
and maximum Extractable Ergotropy (green dashed) for a qubit in the ground state as a function of $T_{\rm E}$.
The work extracted {via the considered bang-bang protocols (red curves)} represents a lower bound for ELE.
We also report GE values (black dotted).
We have set $k_{\rm B}=1$ and
$\omega_{\rm S}= \omega_{\rm E}=1$, $\Omega=0.1$.
Each $U_{\rm bb}(T_{\rm S},T_{\rm E})$ requires less than $100$ local operations.
\label{Fig.2}
}
\end{figure}

For the considered JC model, we show in Fig.\,\ref{Fig.2} that ELE exceeds for large intervals of temperatures the extracted ergotropy even for simple sub-optimal protocols.
We plot as a function of the qubit temperature $T_{\rm S}$ the work extracted with an appropriate unitary operation $ U_{\rm bb}(T_{\rm S},T_{\rm E}) \in \bar{\mathcal{U}}_{\rm ex}$ (with fixed cavity temperature $T_{\rm E} $). In the inset we plot the work extracted by $ U_{\rm bb}(0,T_{\rm E}) \in \bar{\mathcal{U}}_{\rm ex}$ as a function of the cavity temperature $T_{\rm E}$, while $T_{\rm S}=0$. 
The unitaries $U_{\rm bb}(T_{\rm S},T_{\rm E})$ of Fig.\,(\ref{Fig.2}) where obtained 
via a bang-bang protocol of the form \eqref{bang-bang}. 
%
Here time steps $\delta t_k$ are aimed to maximize the ergotropy $\mathcal{E}(\rho_{\rm S}, H_{\rm S})$ of the subsystem S via free SE evolution given by $U_0(\delta t_k)$. Local terms $U_{\rm S}^{(k)}$ implement either the corresponding local unitary that maximizes (local) work extraction or a random energy-preserving local unitary. The latter is implemented only in case free evolution did not act (or equivalently acted with a $\delta t_k=0$) because it could not enhance the ergotropy of the subsystem S on the given input state \cite{supp-mat}.
We emphasize that unitaries $U_{\rm bb}(T_{\rm S},T_{\rm E})$ are not optimal, meaning that the red lines in Fig.\,\ref{Fig.2} represent a lower bound for ${\mathcal{E}}_{\rm ex}(\gamma_{\beta_{\rm S}} \otimes \gamma_{\beta_{\rm E}}) $.
In this setting, GE (and hence ELE) becomes zero when the system and the bath reach the same temperature.
This comes as a consequence of the weak-coupling assumption and from the passivity of the compound thermal state 
with respect to the non-interacting Hamiltonian $H_{\rm S}+H_{\rm E}$.
At lower temperatures $T_{\rm S}$, 
the main plot in Fig.\,\ref{Fig.2} shows how ELE is greater than EE for a considerable temperature interval.
Furthermore, we notice that in the assumed weak-coupling regime, we can define work and heat as follows.
Given a certain protocol for energy extraction, i.e. given a $U \in  \bar {\mathcal{U}}_{\rm ex} $, the work extracted is
$
W= {\rm tr}[H_{\rm SE}(\rho_{\rm in}-
U\rho_{\rm in}U^\dag)]\,
$
and the heat is the energy difference of the thermal bath
$
Q={\rm tr}[H_{\rm E}(\rho_{\rm in}-U\rho_{\rm in} U^\dag)] 
\,.
$
Protocols implemented in Fig.\,\ref{Fig.2} are also characterized by considerably high $
{W}/{Q} 
$ ratio  (in our case $W$ and $Q$ are both positive). 
Indeed, the final local state $ \rho_{\rm out, S}:={\rm tr}_{\rm E}(U\rho_{\rm in} U^\dag) $ is a (completely) passive state, which implies $ {\rm tr}(\rho_{\rm out ,S}H_{\rm S})\leq \omega_{\rm S}/2 $.
This means that the heat exchanged is always smaller than $W+\omega_{\rm S}/2$. For example, we get $W/Q \geq 0.763 $ for a cavity temperature corresponding to $12$ photons on average and the qubit initially in the ground state (last point in the inset of Fig.\,\ref{Fig.2}).

{\it Discussion.---}
We have investigated maximal work extraction from a system interacting with an environment via manipulation of the local Hamiltonian on S. 
We introduced an extended version of local ergotropy 
that exploits the free evolution of the SE compound.
We showed that extended local ergotropy has conceptual and practical advantages with respect to local ergotropy \cite{LErgo} and ergotropy extraction via thermal operations \cite{ThermalOp}.
Most notably, it is not increasing under free time evolution and rules out arbitrary control of the environment or its interaction with the system. 
We then quantified the performances and efficiency of extended local ergotropy in the Jaynes-Cummings model, using both analytical tools from quantum control theory and numerical calculations on practical sub-optimal protocols.

Finally, we notice that our work leaves open questions addressable via quantum control methods, such as identifying the shortest amount of time or number of local unitaries needed to perform the optimal protocol or  
those Hamiltonians for which ELE saturates the Global ergotropy. 

{\it Acknowledgments.---}
We thank Leonardo Zambrano, Raffaele Salvia, Luca Milanese, and Martino Barbieri for useful discussions.
This work was supported by the Government of Spain (Severo Ochoa CEX2019-000910-S and FUNQIP), Fundació Cellex, Fundació Mir-Puig, Generalitat de Catalunya (CERCA program), the AXA Chair in Quantum Information Science, the ERC AdG CERQUTE and the PNRR MUR Project No. PE0000023-NQSTI.
\bibliography{biblio}

\begin{thebibliography}{34}
\bibitem [{Erg()}]{ErgoFormula}%
  \BibitemOpen
  \href@noop {} {\bibinfo  {journal} {Let $\rho= \sum_{j} r_{j} \ket{r_{j}}\bra{r_{j}} $, with $r_{j+1}\leq r_{j}$, and $H=\sum_{j} E_{j} \ket{E_j}\bra{E_j} $, with $E_{j+1} \geq E_{j}$, be the spectral decompositions of the state and Hamiltonian, then the optimal unitary is $ \sum_{j}\ket{E_j}\bra{r_{j}} $}\ }\BibitemShut {NoStop}%
\bibitem [{Clo()}]{Closure}%
  \BibitemOpen
  \href@noop {} {\bibinfo  {journal} {The topological closure of a set X, here denoted as $ \bar X $, is defined as the smallest closed set that contains X. In our case it means that the unitaries that can be approximated arbitrarily well by unitaries of $\mathcal{U}_{\rm ex}$ belong to $ \bar{\mathcal{U}}_{\rm ex}$. For instance, local unitaries are part of the $ \bar{\mathcal{U}}_{\rm ex}$. Indeed, if $U_{\rm S}=e^{-i\tau H'_{\rm S}}= \underset{n \rightarrow \infty}{\lim}\mathcal{T} \exp [-i\int_{0}^{\tau/n} \!\! H_{\rm SE}+nH'_{\rm S}dt ] $}\ }\BibitemShut {NoStop}%
\bibitem [{cyc()}]{cyclic}%
  \BibitemOpen
\bibfield  {journal} {  }\href@noop {} {\bibinfo  {journal} {It is worth mentioning that in a real experiment, Hamiltonian driving should also respect the cyclic condition $H_{\rm S}(0)=H_{\rm S}(t_f)=0$. However, this constraint is irrelevant to the mathematical definition of $\mathcal{U}_{\rm ex} $}\ }\BibitemShut {NoStop}%
\bibitem [{Alm()}]{AlmostsameGE}%
  \BibitemOpen
  \href@noop {} {\bibinfo  {journal} {For the coupling streght used to generate Fig.(1), $\Omega=0.1$ we have ${\mathcal{E}}(\ket{0}_{\rm S}\otimes\ket{12}_{\rm E})={\mathcal{E}}(\ket{0,\alpha=\sqrt{12}}_{\rm SE}) = 12$ and ${\mathcal{E}}(\ket{11+}_{\rm SE})= 12+\sqrt{3}/10 $}\ }\BibitemShut {NoStop}%
\bibitem [{LE=()}]{LE=0}%
  \BibitemOpen
\bibfield  {journal} {  }\href@noop {} {\bibinfo  {journal} {For factorized states one can prove \cite{LErgo} that $\mathcal{E}_{\rm S}(\rho_{\rm S} \otimes \rho_{\rm E},H_{\rm S}+H_{\rm E}+V_{\rm SE})=\mathcal{E}(\rho_{\rm S},H_{\rm S}') $, where $H_{\rm S}'=H_{\rm S}+{\rm tr}_{\rm E}[V_{\rm SE}\rho_{\rm E}] $. In our case the interaction is linear in creation and destruction operators, therefore ${\rm tr}_{\rm E}[V_{\rm SE}\gamma_{\beta_{\rm E}}]=0$, leading to $ \mathcal{E}_{\rm S}(\gamma_{\beta_{\rm S}} \otimes \gamma_{\beta_{\rm E}},H_{\rm S} +H_{\rm E}+V_{\rm SE})=\mathcal{E}(\gamma_{\beta_{\rm S}},H_{\rm S})=0$}\ }\BibitemShut {NoStop}%
\bibitem [{Int()}]{InteractionStrengh}%
  \BibitemOpen
\bibfield  {journal} {  }\href@noop {} {\bibinfo  {journal} {The dynamical Lie algebra $\mathcal{L}$ generated by $\lambda V_{\rm AB} $ and local Hamiltonians on S is independent of $ \lambda $ as long as $\lambda \neq 0 $, see \cite{supp-mat}}\ }\BibitemShut {NoStop}%
\bibitem [{Loc()}]{LocallyBoundedMesurable}%
  \BibitemOpen
\bibfield  {journal} {  }\href@noop {} {\bibinfo  {journal} {Meaning that are measurable w.r.t. Lebesgue measure and $\forall \; t \; \exists $ a neighborhood $I$ of $t$ and a constant $M$ s.t. $ \mu(t)\leq M \;\; \forall t \in I$}\ }\BibitemShut {NoStop}%
\bibitem [{Com()}]{CompactCheckNote}%
  \BibitemOpen
\bibfield  {journal} {  }\href@noop {} {\bibinfo  {journal} {$\mathcal{L}$ being semi-simple, or the Killing form of $\mathcal{L}$ being negative are examples of sufficient conditions. A detailed discussion can be found in section 3.2 of \cite{Q.ControlBook}}\ }\BibitemShut {NoStop}%
\end{thebibliography}

\clearpage

\appendix 

\section{Supplemental material}

\section{ $\bar {\mathcal{U}}_{\rm ex}(H_{\rm SE}) $ in a Quantum Control Theory formalism} 

\label{ControlRevAppendix}
Let $ d_{\rm S/E}$ be the (finite) Hilbert space dimension of S/E.
Then in equation \eqref{LocalOp} of the main text, we can always write $H_{\rm S}(t):=\sum_{j=1}^{d_{\rm S}^2} B_{\rm S}^{(j)}\mu_{j}(t)$, where $B_{\rm S}^{(j)}$ are Hermitian operators that form a basis for the Lie algebra of Hermitian operators acting on the Hilbert space $\mathcal{H}_{\rm S}$. 
In literature, different constraints on the $\mu_{j}(t)$ functions have been imposed.  
To fix ideas, one can assume the $\mu_{j}(t)$ to be measurable and locally bounded \cite{LocallyBoundedMesurable}. However, it is by no means necessary to be able to implement such a vast class of functions, instead even 
allowing them to take only 2 distinct values is enough to generate any desired $U \in {\mathcal{U}}_{\rm ex}~$ \cite{Q.Control(1972)}.
An important conceptual consequence is that at variance with the framework of LE, whenever we use a $U \in \mathcal{U}_{\rm S}\subset \bar{\mathcal{U}}_{ex}$ there is no adiabatic approximation implicitly assumed, local unitaries can be obtained arbitrarily precisely without control on a much shorter time scale than the joint evolution.\\
Let $\mathcal{L}$ be the Lie algebra generated by the discrete set $ \big \{-iB_{\rm S}^{(1)}, -iB_{\rm S}^{(2)}...-iB_{\rm S}^{(d_{\rm S}^2)}, -iH_{SE} \big \}$. 
Procedures to obtain $\mathcal{L}$ are detailed for example in Ref.\,\cite{Q.ControlBook}. Intuitively, they consist in iteratively adding to ${\rm SPAN }\big \{-iB_{\rm S}^{(1)}, -iB_{\rm S}^{(2)}...-iB_{\rm S}^{(d_{\rm S}^2)}, -iH_{SE} \big\} $
all the operators (and their real linear combination) that can be generated by the commutator of pairs of elements in the vector space.
It can be proven that \cite{Q.Control(1972),Q.ControlBook,Q.ControlRev}
\begin{equation}
  \bar {\mathcal{U}}_{\rm ex}=  \big\{ e^{L_{1}} e^{L_{2}}...e^{L_{k}} \; s.t.\; L_{i} \in \mathcal{L} \big \}:=e^{\mathcal{L}}\,.
\end{equation}
This result gives considerable physical insights. If ${\rm dim}(\mathcal{L})\geq d_{E}d_{S}-1 $, we can immediately conclude that $\bar {\mathcal{U}}_{\rm ex} \supseteq SU(d_{S}d_{E}) $ and thus any unitary operation on the joint system can be performed with local controls on S, meaning also that ELE always equals the ergotropy on the joint system.
In Quantum Control literature this property has been called indirect controllability \cite{IndirectControl}. Conditions on the interactions that enable indirect control have been studied in small-finite dimensional systems~\cite{AlgConditionsIndirectControll}, however, they usually do not apply to realistic models of system-environment interactions.

Furthermore, if $ e^{\mathcal{L}} $ is closed \cite{CompactCheckNote}, $\mathcal{U}_{\rm ex} $ is compact and there exists a universal time $\tau(H_{\rm SE})$ and a number $ N_{\rm max}(H_{\rm SE}) $ s.t. every $U \in \mathcal{U}_{\rm ex} $ can be implemented with a bang-bang protocol in a total time $t_{f} \leq \tau(H_{\rm SE})$ and with less than $ N_{\rm max}(H_{\rm SE}) $ operations. 
Notice that this also gives a lower bound on the power of the optimal work extraction protocol.

\section{Controllability in the Jaynes-Cummings model } \label{J-C ControllabAppendix}
In this section, we show for the Jaynes-Cumming model that local Hamiltonian control over the qubit suffices to have approximate density matrix controllability, i.e. any initial state $\rho_{\rm in} $ on the joint system can be driven arbitrarily close to any other unitarily equivalent density matrix $ \rho_{\rm out}$.
Unfortunately, because  $H_{\rm JC}$ is unbounded, two density matrices $ \epsilon $-distant can have an arbitrarily large difference in mean energy.
Hence, in general, approximate controllability is not enough to prove $\mathcal{E}_{\rm ex}=\mathcal{E}$ on the full system.

{To simplify the treatment, in what follows we first consider a low-energy finite-dimensional approximation of the JC model. The approximate density matrix controllability on the exact system will follow taking a limit.}
Consider for each $N \geq 1$ the $2(N+1)$ dimensional vector space  $ \mathbf{S}_{N}:={\rm SPAN}\big \{ \ket{0,0}_{\rm SE},\ket{1,0}_{\rm SE},...\ket{0,N}_{\rm SE},\ket{1,N}_{\rm SE}\big \}.$ 
Let $\Pi_{N} $ be the projector on $\mathbf{S}_{N}$.
Notice that the restriction of the Hamiltonian to this subspace has the same eigenvalues and eigenvectors except for the last one, more precisely:
 \begin{eqnarray*}
     && H_{\rm SE}^{(N)}:=\Pi_{N}H_{\rm SE}\Pi_{N}=\sum\limits_{j=0, \nu = \pm}^{N-1} \ket{j\nu}\bra{j\nu} E_{j,\nu}+\\
    && (\cos(\phi_{N})^2E_{N+}+\sin(\phi_{N})^2E_{N-})\ket{1,N}\bra{1,N}  
 \end{eqnarray*}
For large $N$ w.r.t. the number of photons in the initial state, this finite-dimensional version of the JC model represents a good approximation of the ideal system.
We now show that for $N \geq 1$:
 \begin{equation}
     \big\{ U\; s.t.\; U = \mathcal{T} \exp \big [-i\int_{0}^{t_f} \!\!\!H_{\rm SE}^{(N)}+\sigma_{x}u(t)dt \big ]\big \} \supseteq SU(2(N+1))
 \end{equation}
were $u(t)$ is a piecewise constant function.
This implies that we can generate any unitary operation on the $\mathbf{S}_{N}$ sub-space for any finite $N$, which trivially implies 
${\mathcal{E}}_{\rm ex}(\rho_{\rm SE},H_{\rm JC}^{(N)})={\mathcal{E}}(\rho_{\rm SE},H_{\rm JC}^{(N)}) \,$ as stated in the main text.
This property can be verified thanks to the following observations.
First, except for a countable set $\mathcal{X}_{0} $, for any value of the coupling $\Omega$, 
$|E_{n,\nu}-E_{m,\mu}|\neq |E_{l,\nu'}-E_{k,\mu'}|  $ for all non-trivial choices of indices \cite{J-Controllab}.
Secondly,
\begin{eqnarray}
&&
\bra{00}\sigma_x\ket{0+}=\cos(\phi_{0})\,\neq 0,\\
&& \bra{n+1+}\sigma_x\ket{n+}=\sin{\phi_{n}}\cos{\phi_{n+1}}\,\neq 0, \\
&& \bra{n+1+}\sigma_x\ket{n-}=-\cos{\phi_n}\cos{\phi_{n+1}}\,\neq 0,\\
&& \bra{1,N}\sigma_x\ket{N-1+}=\sin(\phi_{N-1})\,\neq 0.
\end{eqnarray}
These two observations guarantee for any $ N \geq 1$ and $\Omega \in \mathbf{R}\setminus\mathcal{X}_{0} $  the existence of a non-resonant connectedness chain \cite{InfDimControllabTest}, proving the claim via proposition 3.1 of the reference.
Letting $ N \rightarrow \infty $, we also obtain approximate density matrix controllability of the full system via theorem 2.11 of \cite{InfDimControllabTest}.

\section{Protocols for work extraction} \label{ProtocolAppendix}
In what follows we present details of the considered protocols for work extraction.
These protocols allow us to identify a "good" unitary operation and a short sequence of local operations and free-time evolutions to implement such unitary. 
\subsection{Thermal work extraction}
The unitary $U_{\rm bb}(T_{\rm S},T_{\rm E})$ of Fig.\,\ref{Fig.2} of the main text was obtained following bang-bang recursive steps based on the knowledge of the ergotropy of the subsystem S.
We initially set $ \rho_{\rm in}=\gamma_{\beta_{\rm S}}\otimes \gamma_{\beta_{\rm E}}\equiv \rho_{0}$ (iteration $k=0$).\\
(a) At iteration $k$, we find a good interaction time $\delta t_k \in [0, \tau]$ as 
\begin{equation}
 \delta t_k:=\underset{\delta t \in [0, \tau]}{\rm argmax}\, {\mathcal{E}}(\rho_{k, \rm S}(\delta t),H_{\rm S})\,,   
\end{equation}
where $\rho_{k, \rm S}(\delta t)={\rm tr}_{\rm E} (U_{0}(\delta t)\rho_{k}U_{0}(\delta t)^{\dagger})$ is the reduced state of the system S.
\\
(b) If $\delta t_k=0$ (meaning that free SE evolution cannot increase the ergotropy of the subsystem S), we apply a random energy preserving local unitary on S, $V_{\rm S}^{(k)}$, 
and we go back to step (a) with the resulting new density matrix $\rho_{k+1}=V_{\rm S}^{(k)} \rho_k {V_{\rm S}^{(k)}}{}^\dag$.\\
Otherwise,
we define the unitary on S 
\begin{equation}
W_{\rm S}^{(k)}:=\underset{U_{\rm S} \in \mathcal{U}_{\rm S}}{\rm argmax}\;\; 
{\rm tr}\{ H_{\rm S}
[\rho_{k,\rm S}(\delta t_k)-U_{\rm S}\rho_{k,\rm S}(\delta t_k)U_{\rm S}^\dag ]\}    
\end{equation}
and apply it to the evolved state:
$ \rho_{k+1}=W_{\rm S}^{(k)} \, \rho_{k}(\delta t_{k}) 
\, {W_{\rm S}^{(k)}}^\dag$.
This implies work extraction from the system at iteration $k$.
We then pass to iteration $k+1$ using the new input state $\rho_{k+1}$, going back to step (a).\\
The final form of $U_{\rm bb}(T_{\rm S}, T_{\rm E})$ is a bang-bang unitary of the form \eqref{bang-bang} of the main text, namely 
\begin{eqnarray}
U_{\rm bb}(T_{\rm S}, T_{\rm E}) = 
U_{\rm S}^{(\mathcal{N}-1)} U_{0}(\delta t_{\mathcal{N}-1})
\dots
{U_{\rm S}^{(0)}}U_{0}(\delta t_0)\nonumber
\,,
\end{eqnarray}
with 
$
U_{\rm S}^{(k)}
$
being either a local unitary for work extraction or an energy-preserving random local unitary, symbolically
$
U_{\rm S}^{(k)}
\in \{W_{\rm S}^{(k)},V_{\rm S}^{(k)}\}$.
\\
In Fig.\,\ref{Fig.2} of the main text, the above maximizations have been performed through discretizations of the time intervals $[0,\tau]$, with $\tau=\frac{3\pi}{\Omega}$. The algorithm terminates either when $100$ local unitaries (random or not) are applied or when $7$ random unitaries in a row are applied.
Usually, the latter turns out to be the relevant condition for termination.
Finally, for each point of Fig \ref{Fig.2} of the main text, we repeated the protocol 5 times independently and then chose the maximum work extracted among those. This corresponds to a mildly lucky choice of the random gates.
\subsection{ Work extraction for pure states}
The unitaries used for the work extraction on pure states, i.e. those used to generate the plots of Fig.~\ref{Fig.1} of the main text are obtained with the same procedure and parameters except for the use of random energy preserving unitaries, i.e. in case of (b) the protocol terminates immediately. Notice that up to the numerical error, this protocol equals the unitary in equation \eqref{bb-prot} of the main text if $\rho_{\rm in}=\ket{0,N}\!\bra{0,N} $. It is worth mentioning that the different unitaries used for the work extraction of Fig. \ref{Fig.1} of the main text took different total times of free evolution to be computed, in particular, $\tau(\ket{0,12}) \approx 17.6/\Omega  $,
$\tau(\ket{12+}) \approx 38.4/\Omega $, $\tau(\ket{0,\alpha=\sqrt{12}}) \approx 93.4/\Omega  $, leading to significantly different powers.

\subsection{Explicit computation of Eq.~\eqref{bb-prot} of the main text}
First, we notice that for $\Delta \omega =0 $ we have  $ \phi_{n}=\pi/4, \;\; \forall n \in \mathbf{N}$. Thus, each state with qubit in the ground state and $n+1$ photons in the cavity can be written as $\ket{0,n+1}=\frac{1}{\sqrt{2}}(\ket{n+}-\ket{n-}) $, while the (isoenergetic) state with excited qubit and $ n $ photons in the cavity can be written as $\ket{1,n}=\frac{1}{\sqrt{2}}(\ket{n+}+\ket{n-}) $. Since $\ket{n+} $ and $\ket{n-}$ are eigenvectors split in energy by $2 \Delta \omega_{n} $, this $ e^{i\pi}$ relative phase change can be induced by a free time evolution lasting $\delta t_{n}:=\frac{\pi}{2\Delta \omega_{n}} $, that is, 
$U_{0}(\delta t_{n})\ket{0,n+1}=\ket{1,n} $. We can now extract the energy from the qubit with a (instantaneous) bit flip on the qubit, $U_{\rm S}^{(\rm bf)} U_{0}(\delta t_{n})\ket{0,n+1}=\ket{0,n}$. Eq.~\eqref{bb-prot} of the main text is then obtained by applying the same argument to $\ket{0,n} $, and so on.\\ 
The total time of free time evolution employed to extract the $ n+1 $ photons from the cavity is then: 
\begin{equation}
    T_{n}^{\rm (tot)}=\sum_{j=0}^{n} \delta t_{j}= \frac{\pi}{\Omega} \sum_{j=0}^{n} \frac{1}{\sqrt{j+1}}
\end{equation}
Interestingly, the power of the protocol grows asymptotically as $\sqrt{n}$.

\section{Time dependence of ELE and proof of Eq.~\eqref{constant} of the main text} \label{Time-DependeceAppendix}

Through the proof, let $\rho_{\rm SE}(t) $ be the state of the joint system evolved under $ H_{\rm SE}$.
 We use two known lemmas.\\
{\it Lemma 1}:\\
If $H_{\rm SE}$ has discrete spectrum \cite{Noteonrecurr, Q.Recurr, Recurr}:
\begin{equation}
\forall \epsilon, T > 0 \;\; \exists \;\;t_{\epsilon}>T \;\text{s.t.} \; D(\rho_{\rm SE}(t_{\epsilon}),\rho_{\rm SE}(0)) \leq \epsilon\,.
\end{equation} 
{\it Lemma 2} (continuity of the energy functional):
\begin{equation}
|{\rm tr}[(\rho-\sigma)H_{\rm SE}]|\leq 2D(\rho,\sigma)\Vert H_{\rm SE}\Vert_{\infty}\,,
\end{equation}
where
$D(\rho, \sigma)$ is the trace distance and $\Vert H_{\rm SE}\Vert_{\infty}$ is the energy gap between the lowest and highest Hamiltonian eigenvalues.

Let $U_{n} \in \mathcal{U}_{ex} \; \forall n \in \mathbf{N} $ be a sequence of unitaries such that:
\begin{equation}
    \underset{n \rightarrow +\infty}{{\rm lim}} {\rm tr}[H_{\rm SE}(\rho_{\rm SE}(0)-U_{n}\rho_{\rm SE}(0)U_{n}^{\dagger})]={\mathcal{E}}_{\rm ex}(\rho_{\rm SE},H_{\rm SE}).
\end{equation} 
Such sequence must exist by the definition of ELE.
Since the trace distance is invariant under unitary operations, for any $\epsilon > 0 \;\; \exists \; t_{\epsilon}  $ such that $\forall n \in \mathbf{N}$ :
\begin{equation}
    D(U_{n}\rho_{\rm SE}(0)U_{n}^{\dagger}),U_{n}\rho_{\rm SE}(t_{\epsilon}+T)U_{n}^{\dagger}))\leq \epsilon\,.
\end{equation}
Thus, using {\it Lemma 2} we get for any $\epsilon > 0$ and $\forall n \in \mathbf{N}$ 
\begin{align*}
    & \big |{\rm tr}\big [ H_{\rm SE}\big ( U_{n}(\rho_{\rm SE}(0)-\rho_{\rm SE}(t_{\epsilon}+T))U_{n}^{\dagger})\big ] \big |
     \nonumber
     \\ 
    &  \leq 2 \epsilon \Vert H_{\rm SE}\Vert_{\infty},
\end{align*}
which implies
\begin{equation}
    |\mathcal{E}_{\rm ex}(\rho_{\rm SE}(t_\epsilon),H_{\rm SE})-\mathcal{E}_{\rm ex}(\rho_{\rm SE}(0),H_{\rm SE})| \leq 2 \epsilon \Vert H_{\rm SE}\Vert_{\infty}\,,
 \end{equation}
leading to the thesis.

We now sketch an example of systems with continuous and unbounded Hamiltonian, thus violating the hypothesis of Eq. \!(\ref{constant}) of the main text, in which the energy irreversibly flows into the environment, making ELE strictly decrease in time. 
Consider a neutral particle with spin 1/2 (for example a neutron) constrained in 1D and immersed in a magnetic field of the form $\vec B(x)={\rm sgn(x)}B_{0}\hat{{\rm z}} $. Its Hamiltonian is 
\begin{equation}
    H_{\rm SE}^{B}=\frac{P^{2}}{2m} -\mu \sigma_{z}{\rm sgn(x)}B_{0}
\end{equation} 
where we have associated the spin degrees of freedom to the controllable system S and the spatial degrees of freedom to the environment E. 
Notice that $[H_{\rm SE}^{B},\sigma_{z}]=0 $ , so its eigenvectors can be chosen to be separable.
In particular, for $k\geq 0 $ let $k_{+}(k)=\sqrt{2m(k^2+4\mu B_{0})} $ and let the transmission and reflection amplitude be respectively $t(k)=\frac{2k}{k+k_{+}}$, $r(k)=t(k)-1$. Then
we have 
\begin{equation}
    H_{\rm SE}^{B}\ket{\downarrow} \ket{\psi(k)}=\left(\frac{k^2}{2m}+\mu B_{0} \right) \ket{\downarrow}\ket{\psi(k)}
\end{equation}
where 
\begin{eqnarray}
 && \braket{x|\psi(k)}:={\rm e}^{ikx} +r{\rm e}^{-ikx} \;\;{\rm if}\;\; x\leq 0 \\
 && \braket{x|\psi(k)}:=t{\rm e}^{ik_{+}x} \;\; {\rm if} \;\;x > 0\,.
\end{eqnarray}
Further, let $T=|t|^{2}$, $R=|r|^{2} $ be the transmission and reflection probabilities.\\
Now consider as initial state a spin-down wave packet with support in the $x< 0$ region of space and moving towards the right with group velocity $v_{g}=\bar{k}/m $, where $\bar{k} $ is chosen s.t. $T(\bar k)=|t(\bar k)|^2=1/2$. At time $t=0$ the state has an ELE of at least $2\mu B_{0} $ due to the Zeeman splitting. A simple bit flip of the spin can extract such work.
After a sufficient time has passed, the particle has scattered against the abrupt magnetic field change, and the wave packet has a split in two, one traveling towards the left with $-\bar{k}$ momentum, the other towards the right with $k_{+}(\bar{k})$ momentum. Both wave packets still have spin-down and their integral of the square modulus is equal to 1/2 each. 
Since both packets are now far away from the inhomogeneity of the magnetic field, local operations on the spin will not result in any change in the spatial degrees of freedom. 
It is immediate to verify that any local operation on the spin does not change the total energy of the system because the wave functions have equal weight on regions of space with opposite magnetic fields. We conclude that ELE is now 0, the energy initially stored in the exited spin state has irreversibly flown into the kinetic energy of the transmitted wave packet.

\section{Spin chain model}
 We showed that in the Jaynes-Cummings model, except for special pure initial states, it is not possible to extract all work from the compound system.  
Now we apply the concept of ELE to two spin models, showing that in some physical models, the environment can be instead efficiently controlled, even if arbitrarily large.  
Consider a 1D chain of M 1/2 spins with Heisenberg-like first neighbors interaction, 
\begin{equation}
   H_{SE}=\underset{n=1}{\overset{M-1}{\sum}} \gamma (\sigma^{(n)}_{x}\sigma^{(n+1)}_{x}+\sigma^{(n)}_{y}\sigma^{(n+1)}_{y}+\Delta \sigma^{(n)}_{z}\sigma^{(n+1)}_{z}).
\end{equation}
We consider as the $S$ system only the first spin from the left, spin 1, and as the environment all other spins $2, 3, \dots, M$. 
Correspondingly, the set of extended local operations can be written as: 
\begin{equation} 
    \mathcal{U}_{\rm ex}(H_{\rm SE})=\big\{ U\; s.t.\; U = \mathcal{T} \exp \big [-i\int_{0}^{t_f} \!\!\! H_{\rm SE}+\vec{\sigma}^{(1)}\cdot \vec{V}(t)dt \big ]\big \} \,.
\end{equation}

Physically, local Hamiltonian control $ \vec{\sigma}^{(1)}\cdot \vec{V}(t) $ can conveniently be implemented with a time-dependent magnetic field localized on the first site of the chain. In \cite{LocalControlSpinChain(General)}, they were able to prove that if \(\Delta \neq 0\), the dynamical Lie algebra generated from \(H_{\rm SE}\) and \(\vec{\sigma}^{(1)}\) equals the whole Lie algebra of self-adjoint operators on the \(M\) qubits, meaning that \(\mathcal{U}_{\rm ex}(H_{\rm SE}) \supseteq SU(2^{M})\). In other words, by only manipulating the magnetic field of the first spin any unitary operation can be carried out on the whole spin chain.
Consequently, for any state, we have 
\begin{equation}
 {\mathcal{E}}_{\rm ex}(\rho_{\rm SE},H_{\rm SE})={\mathcal{E}}(\rho_{\rm SE},H_{\rm SE}) \,.
\end{equation} 
In this case, $\mathcal{U}_{ex}$ is compact and this guarantees us that all $U\in \mathcal{U}_{ex}$ can be perfectly implemented within a universal time $\tau(H_{\rm SE})$. However, we do not possess estimates on how large $\tau(H_{\rm SE})$ is and in particular how it scales with $M$. An interesting possibility to overcome this difficulty is considering the case $\Delta=0$, where it was proven in \cite{ControlSpinChain-2spins} that control on the first 2 spins is enough to \textit{efficiently} control the whole chain. In particular, implementing a gate takes quadratically longer than for direct control.
This example highlights that our framework can also be seen from another perspective: the whole spin chain can be seen as the work storage device, and the first two spins are the \lq\lq plug" thanks to which we can extract all work in a practical way.

\end{document}